\documentclass[prd,onecolumn,showpacs,nofootinbib]{revtex4-1}
\usepackage{amsmath}
\usepackage{latexsym}
\usepackage{amsfonts}
\usepackage{graphicx}
\usepackage{mathrsfs}
\usepackage{hyperref}
\usepackage{mathrsfs}
\usepackage{accents}
\usepackage{color}

 \newcommand{\bq}{\begin{equation}}
 \newcommand{\eq}{\end{equation}}
 \newcommand{\bqn}{\begin{eqnarray}}
 \newcommand{\eqn}{\end{eqnarray}}
 \newcommand{\nb}{\nonumber}
 
\begin{document}

\title{Waveforms of compact binary inspiral gravitational radiation in screened modified gravity}

\author{Tan Liu$^{1,2}$}\email{lewton@mail.ustc.edu.cn}
\author{Xing Zhang$^{1,2}$}\email{starzhx@mail.ustc.edu.cn}
\author{Wen Zhao$^{1,2}$}\email{wzhao7@ustc.edu.cn}
\author{Kai Lin$^{3,4}$}
\author{Chao Zhang$^5$}
\author{Shaojun Zhang$^3$}
\author{Xiang Zhao$^5$}
\author{Tao Zhu$^3$}
\author{Anzhong Wang$^{3,5}$}\email{Anzhong$_$Wang@baylor.edu}

\affiliation{$^1$CAS Key Laboratory for Researches in Galaxies and Cosmology, Department of Astronomy, \\
University of Science and Technology of China, Chinese Academy of Sciences, Hefei, Anhui 230026, China}
\affiliation{$^2$School of Astronomy and Space Science, University of Science and Technology of China, Hefei 230026, China}
\affiliation{$^3$Institute for Advanced Physics $\&$ Mathematics, Zhejiang University of Technology, Hangzhou 310032, China}
\affiliation{$^4$Universidade Federal de Itajub\'a, Instituto de F\'isica e Qu\'imica, Itajub\'a, MG, Brasil}
\affiliation{$^5$GCAP-CASPER, Physics Department, Baylor University, Waco, TX 76798-7316, USA}

\begin{abstract}
Scalar-tensor gravity, with the screening mechanisms to avoid the severe constraints  of the fifth force in the Solar System, can be described
with a unified theoretical framework, the so-called screened modified gravity (SMG). Within this framework,   in this paper we calculate the
waveforms of gravitational-waves (GWs) emitted by inspiral compact binaries, which include four polarization modes, the plus $h_{+}$, cross
$h_{\times}$, breathing $h_{b}$, and  longitudinal  $h_{L}$ modes. The scalar polarizations $h_b$ and $h_L$ are both caused by the scalar
field of SMG, and satisfy a simple linear relation. With the stationary phase approximations, we obtain their Fourier transforms, and derive the
correction terms in the amplitude, phase, and polarizations of GWs, relative to the corresponding results in general relativity. The corresponding
parametrized post-Einsteinian parameters in the general SMG are also identified. Imposing the noise level of the ground-based Einstein Telescope,
we find that GW detections from inspiral compact binaries composed of a neutron star and a black hole can place stringent constraints on the
sensitivities of neutron stars, and the bound is applicable to any SMG theory. Finally, we apply these results to some specific theories of SMG,
including chameleon, symmetron, dilaton and $f(R)$.

\end{abstract}


\maketitle

\section{Introduction}

Einstein has laid the foundation of general relativity (GR) \cite{Einstein1916} and gravitational waves (GWs) \cite{Einstein1918} more than one hundred years ago. In recent years, the LIGO  and Virgo collaborations have detected several GWs from
binary systems, and realized our century-long dreams of detecting GWs directly \cite{PhysRevLett.116.061102,PhysRevLett.116.241103,PhysRevLett.118.221101,Abbott2017a,Abbott2017,2041-8205-851-2-L35}. This inaugurates the new era of  gravitational astronomy. Since GR was proposed, it has been tested in various circumstances  \cite{will1993theory,Will2014}. However, most of these tests focused mainly on the weak field regimes. The coalescence of a compact binary system can produce strong gravitational fields. Therefore, the GW observations allow us to test GR in the highly dynamical and strong field regime for the first time \cite{thorne1987gravitational}.

It is well known that there exist two independent GW polarizations $h_+$ and $h_\times$ in GR \cite{maggiore2008gravitational,misner1973gravitation}. However, in a metric theory of gravity, considering the symmetric properties of the Riemann tensor and the Bianchi identity, there can be at most six different polarizations \cite{will1993theory}.  Eardley and collaborators developed the $E(2)$ classification scheme of GW polarizations to classify metric theories of gravity, but their discussions are limited to null GWs \cite{PhysRevLett.30.884,PhysRevD.8.3308}. This scheme is based on the transformation properties of the polarizations under the little group $E(2)$ of the Lorentz group. Afterwards, the $E(2)$ classification scheme is extended to include nearly all null waves in \cite{will1993theory}. A GW detector measures a linear combination of the GW polarizations, which is called the response function \cite{maggiore2008gravitational}.

With GW detections, we can test GR in two different approaches, one is  theory-independent and the other is theory-dependent.  In the theory-independent approach, the deviations from GR are characterized by several parameters. Theory-independent tests can constrain many different theories at the same time. The parametrized post-Einsteinian (ppE) framework is a  theory-independent approach. The standard ppE framework was proposed by Yunes and Pretorius \cite{PhysRevD.80.122003}, and they only considered the two tensor polarizations, $h_+$ and $h_\times$, emitted by a compact binary on a quasicircular orbit. The Fourier transform of  the response function in metric theories of gravity is parametrized by four ppE parameters in the standard ppE framework.  Recently, the standard ppE framework has been extended to include all the six polarizations and there are more parameters in this extended ppE framework \cite{Yunes-BD}. In contrast to the theory-independent approach, the theory-dependent approach constrain a specific  theory by comparing GW waveforms of this theory with GW signals. Although this approach can only test one particular theory at a time, it can directly constrain the fundamental physics in this theory.

In this paper, we construct the GW response function in the screened modified gravity (SMG) for theory-dependent tests of GR. We only consider compact binaries on quasicircular orbits, as the radiation reaction can circularize the orbit to a great accuracy \cite{PhysRev.136.B1224}.  SMG is a scalar-tensor theory with screening mechanisms and is a simple extension of GR. In SMG there are a conformal coupling function $A(\phi)$ and a scalar potential  $V(\phi)$. The scalar potential can act as dark energy to accelerate the expansion of the Universe. The behavior of the scalar field is controlled by an effective potential, which is defined through $V(\phi)$ and   $A(\phi)$ and depends on the environmental density. The fluctuation about the minimum of  the effective potential acquires an environmental dependent mass $m_s$, which is an increasing function of  the local matter density. Then, the scalar field can be screened in high density regions due to the short range of the fifth force \cite{Burrage2018}.

As  natural extensions of GR, scalar-tensor theories have been studied for decades \cite{PhysRev.124.925,faraoni2004cosmology,1975ApJ...196L..59E,Zhang2016,PhysRevD.62.024004,PhysRevLett.108.081103,PhysRevD.70.123518,1989ApJ...346..366W,PhysRevLett.83.2699,PhysRevLett.109.241301,0264-9381-29-23-232002,PhysRevD.85.102003,PhysRevD.87.104029,PhysRevD.97.064016,PhysRevD.89.084005}. The leading order GW waveforms produced by binary systems in Brans-Dicke theory have been calculated in \cite{PhysRevD.50.6058}. These calculations were extended to higher post-Newtonian (PN) orders in \cite{Will-BDn,PhysRevD.94.084003,PhysRevD.98.044004}. In these works, the authors ignored the breathing polarization $h_b$ produced by the scalar field. The breathing polarization $h_b$  in Brans-Dicke theory was obtained in \cite{Yunes-BD,PhysRevD.95.124008,Lang-BD}. However, all these works focused on the scalar-tensor theory with massless scalar field. The GW energy flux in the massive Brans-Dicke theory was worked out in \cite{PhysRevD.85.064041}, but the screening mechanism was not adopted. In \cite{Zhang2017}, taking into account the screening mechanism in SMG, we obtained the GW energy flux emitted by the compact binary system, as well as the  solutions of the tensor  and  scalar fields which are expressed in terms of the mass quadrupole moment and the scalar multipole moments, respectively.

In this paper, based on the results of \cite{Zhang2017}, we work out in details the GW waveforms produced by an inspiral compact binary system on a quasicircular orbit in SMG. We find that there are four polarizations in SMG, i.e., the plus polarization $h_+$, the cross polarization $h_\times$, the breathing polarization $h_b$ and the longitudinal polarization $h_L$. In addition, there is a simple linear relation between $h_b$ and $h_L$ stemming from the scalar field equation, and only three dynamical degrees of freedom exist in SMG. The relation between $h_b$ and $h_L$ is consistent with the previous result \cite{PhysRevD.62.024004}.
In the original $E(2)$ classification, the authors pointed out that for a given theory, if the degrees of freedom of the gravitational field is less than the number of polarizations, these polarizations are linearly dependent in a manner dictated by the detailed structure of the theory \cite{PhysRevD.8.3308}. The relation between $h_b$ and $h_L$ is a good example of this statement. Using the stationary phase approximation, we derive the Fourier transforms of the GW waveforms. Comparing with the predictions in GR, we identify the four ppE parameters of SMG. Then,  we forecast the constraints that the Einstein Telescope may impose on SMG. Applying these constraints to some specific SMG models, including chameleon model \cite{PhysRevLett.93.171104,PhysRevD.69.044026}, symmetron model \cite{PhysRevLett.104.231301}, and dilaton model \cite{PhysRevD.82.063519}, we obtain constraints on the model parameters.

It is well known that $f(R)$ gravity can be rewritten as a scalar-tensor theory \cite{RevModPhys.82.451,LIU2018286,NOJIRI201159,NOJIRI20171}. Therefore, our results of SMG can be applied to $f(R)$ gravity, too. In doing so, we obtain the GW waveforms produced by an inspiral compact binary system in the general $f(R)$ gravity with screened mechanisms, and derive the ppE parameters of $f(R)$ theory.\footnote{Note that, the number of degree of freedom of GW in general $f(R)$ theory is also derived in \cite{Gong-fR} }. Then,  we constrain three specific $f(R)$ models, including the Starobinsky model \cite{starobinsky2007disappearing}, Hu-Sawicki
model \cite{Hu:2007nk} and Tsujikawa model \cite{PhysRevD.77.023507}.

The rest of the paper is organized as follows: In Sec. \ref{smg}, we briefly review SMG. In
Sec. \ref{radiation}, we investigate the orbital motion of the compact binary system and the orbital decay driven by the gravitational radiation. In Sec. \ref{waveform},  we calculate the GW waveforms and their Fourier transforms in SMG. In Sec. \ref{ppe}, we calculate the ppE parameters in SMG and constrain three specific SMG models. In Sec. \ref{frgw}, we apply the results of SMG to $f(R)$ gravity, while in Sec. \ref{con},  we summarize our main results and present some concluding remarks.

For the metric, Riemann and Ricci tensors, we follow the conventions of Misner, Thorne and Wheeler \cite{misner1973gravitation}. We set the units so that $c=\hbar=1$, and therefore the reduced Planck mass is $M_\text{Pl}=\sqrt{1/8\pi G}$, where $G$ is the Newtonian gravitational constant.

\section{Screened modified gravity}\label{smg}

SMG is the scalar-tensor theory with screening mechanisms.
The action of a general scalar-tensor theory in the Einstein frame takes the form
\begin{equation}\label{action}
S=\int d^4 x\sqrt{-g}\left[\frac{1}{16\pi G}R-\frac12 \partial_\mu\phi\partial^\mu\phi-V(\phi)\right]+S_m\left[A^2(\phi)g_{\mu\nu},\Psi_m\right],
\end{equation}
where $g_{\mu\nu}$ is the  metric in the Einstein frame, $g$  its determinant, $R$  the Ricci scalar derived from $g_{\mu\nu}$, $\phi$  the scalar field, $V(\phi)$  the potential, and $A(\phi)$  the conformal coupling function. $\Psi_m$ denotes collectively the mater fields. Because of the conformal coupling function $A(\phi)$, there is a direct interaction between the scalar field and the matter fields. Therefore,  the scalar field will generate a fifth force that will be felt by the matter fields. Since there is no evidence of  the fifth force in the Solar System \cite{Will2014}, we need a mechanism to screen  it in the high density environments. The scalar-tensor theory with a screening mechanism is called screened modified gravity. The screening mechanism will be explained in the following section.

For a compact object, its internal gravitational energy contributes to its total mass.
In the scalar-tensor theory, the effective gravitational constant depends on the local value of the scalar field. Thus, the scalar field can affect the internal structure of a compact object and its total mass. Eardley suggested that the constant inertial mass $m$ of the compact object should be replaced by a function of the scalar field $\phi$, i.e., $m(\phi)$ \cite{1975ApJ...196L..59E}. Then the matter action in Eq.\eqref{action} becomes
\begin{equation}
S_m=-\sum_{a=1}^2\int m_a(\phi)d\tau_a{}.
\end{equation}
Variations of the action $S$, respectively, with respect to $g_{\nu\nu}$ and $\phi$  yield the field equations
\begin{equation}\label{h-eq}
G_{\mu\nu}=8 \pi G(T_{\mu\nu}+T_{\mu\nu}^{\phi}),
\end{equation}
and
\begin{equation}\label{phi-eq}
\nabla_\mu\nabla^\mu\phi=\frac{\partial}{\partial\phi}(V(\phi)-T),
\end{equation}
where
\begin{equation}
T^{\mu\nu}=\frac{1}{\sqrt{-g}}\sum_{a=1}^2 m_a(\phi)\frac{u_a^\mu u_a^\nu}{u_a^0}\delta^{(3)}({\bf x}-{\bf x}_a(t)),
\end{equation}
is the energy-momentum tensor of point particles with $u_a^\mu$ the four-velocity of the particle  $a$, and $T=g^{\mu\nu}T_{\mu\nu}$ is the trace of $T_{\mu\nu}$. The energy-momentum tensor of the scalar field is
\begin{equation}
T^{\phi}_{\mu\nu}=\partial_\mu\phi\partial_\nu\phi -\frac12g_{\mu\nu}\left[\partial_\alpha\phi\partial^\alpha\phi+2V(\phi)\right].
\end{equation}

It can be shown that the behavior of the scalar field is controlled by the effective potential
\begin{equation}\label{veff}
V_{\rm eff}(\phi)=V(\phi)-T.
\end{equation}
For a negligibly self-gravitating object, the effective potential can be rewritten as \cite{Zhang2016}
\begin{equation}
V_{\rm eff}(\phi)=V(\phi)+\rho A(\phi),
\end{equation}
where $\rho$ is the conserved energy density in the Einstein frame \cite{PhysRevLett.109.241301}.

\section{Gravitational radiation in SMG}\label{radiation}

It is well known that  there is no mass dipole radiation in GR as a result of the law of conservation of momentum, and quadrupole radiation is the leading order contribution to the gravitational radiation \cite{maggiore2008gravitational,misner1973gravitation}. However, in the scalar-tensor theory, the scalar dipole moment does not vanish in the center-of-inertial-mass frame, and the compact binary system generally exhibits a time-dependent scalar dipole moment \cite{1975ApJ...196L..59E}. Therefore, the scalar dipole radiation exists in the scalar-tensor theory.  In this section, we review some results from \cite{Zhang2017} about the motion and gravitational radiation  of a compact binary system. The details can be found in \cite{Zhang2017}

In the wave zone (faraway from the binary system), the metric tensor and the scalar field can be expanded around the Minkowski background $\eta_{\mu\nu}$ and the scalar background $\phi_0$, respectively,
\begin{equation}
g_{\mu\nu}=\eta_{\mu\nu}+h_{\mu\nu},\quad \phi=\phi_0+\varphi.
\end{equation}
The bare potential $V(\phi)$ and the coupling function $A(\phi)$ can be expanded around $\phi_0$ as follows,
\begin{align}\label{VA_expand}
\begin{split}
V(\phi)
&=V_0+V_1\varphi+V_2\varphi^2+V_3\varphi^3+\mathcal{O}\left(\varphi^4\right)\,,
\\
A(\phi)
&=A_0+A_1\varphi+A_2\varphi^2+A_3\varphi^3+\mathcal{O}\left(\varphi^4\right)\,.
\end{split}
\end{align}
Then,  the effective mass of the scalar field is
\begin{equation}
m^2_s \equiv \frac{\rm d^2 V_{\rm eff}}{\rm d \phi^2 }\Big |_{\phi_0}=2\left(V_2+\rho_b A_2\right)\,.
\end{equation}
Thus, the effective mass of the scalar field $m_s$ depends on the background matter density $\rho_b$. In the high density environment, the mass $m_s$ becomes large and the  range of the fifth force is too short to be detectable by the Solar System experiments. In the low density cosmological background, the magnitude of the scalar mass can be of the Hubble scale to drive the acceleration of the universe \cite{PhysRevD.82.063519}. As a result, the scalar field is screened in high density environments (e.g. the Solar System), while in the low density environments (e.g., the cosmological scales), it plays a crucial role. This is the so-called screening mechanism.

In the weak-field limit, linearizing the field equations \eqref{h-eq} and \eqref{phi-eq} yields \cite{Zhang2017}
\begin{align}\label{linear_tensor_eq}
\square\bar{h}_{\mu\nu}=-16\pi G\tau_{\mu\nu}\,,
\end{align}
and
\begin{align}\label{linear_scalar_eq}
\left(\square-m^2_s\right)\varphi=-16\pi GS\,,
\end{align}
where $\bar{h}_{\mu\nu}=h_{\mu\nu}-\frac12\eta_{\mu\nu}h^{\alpha}_{\alpha}$ is the trace reversed metric perturbation, $\tau_{\mu\nu}$ is the total energy-momentum tensor and $S$ is the source term of the scalar field. The expressions of $\tau_{\mu\nu}$ and $S$ are given by Eqs. (16) and (19) in \cite{Zhang2017}.
The inertial mass of the compact object $m_a(\phi)$ can also be expanded around the scalar background $\phi_0$,
\begin{align}\label{inertial_mass}
\begin{split}
m_a(\phi)=&m_a\Bigg[1+s_a\left(\frac{\varphi}{\phi_0}\right)+\mathcal{O}\left(\frac{\varphi}{\phi_0}\right)^2\Bigg]\,,
\end{split}
\end{align}
where $m_a=m_a(\phi_0)$ and
\begin{equation}
s_a\equiv\frac{\partial(\ln m_a)}{\partial(\ln \phi)}\bigg|_{\phi_0},
\end{equation}
{is the sensitivity, which characterizes how the gravitational binding energy of a compact object responds to its motion relative to the additional fields.
In SMG, the object's sensitivity is proportional to its screened parameter $\epsilon_a$ \cite{Zhang2017},
\begin{equation}
s_a=\frac{\phi_0}{2 M_{\rm Pl}}\epsilon_a~.
\end{equation}
Considering the object (labeled as $a$) with uniform density, the screened parameter (i.e. the scalar charge) has been calculated previously, which is given by \cite{Zhang2017}
\begin{equation}
\epsilon_a=\frac{\phi_0-\phi_a}{M_{\rm Pl}\Phi_a}\,,
\end{equation}
where $\Phi_a=Gm_a/R_a$ is the surface gravitational potential of the a-th object, and $\phi_a$ is the position of the minimum of the effective potential $V_{\rm eff}$ inside this object and is generally inversely correlated to the matter density $\rho$ \cite{Zhang2016}. Since the background matter density is always much less than that of the compact object, we have $\phi_0\gg\phi_{a}$.
}

In a inspiral compact binary system, we treat the compact objects as point particles and denote their masses as $m_1$ and $m_2$ and their positions as $\mathbf{r_1}$ and $\mathbf{r_2}$.
In the center-of-inertial-mass frame, this two-body system can be reduced to a one-body system, i.e., a point particle with reduced mass $\mu=m_1 m_2/(m_1+m_2)$ orbits around the total mass $m=m_1+m_2$. The equation of motion is \cite{Zhang2017}
\begin{equation}
\frac{d^2 \mathbf{r}}{dt^2}=-\frac{\mathcal{G}m\mathbf{r}}{r^3}~,
\end{equation}
where $\mathbf{r}\equiv\mathbf{r_1}-\mathbf{r_2}$ is the relative coordinate, and the effective Newtonian constant $\mathcal{G}$ is given by \cite{Zhang2017}
\begin{equation}
\mathcal{G}=G\left(1+\frac12 \epsilon_1\epsilon_2\right)~.
\end{equation}

During the gravitational radiation of the compact  inspiral system, the orbital eccentricity decreases very quickly, and the orbital eccentricity is expected to be essentially zero before the binary enters the frequency bandwidth of ground-based GW detectors \cite{PhysRev.136.B1224}. For this reason, in this paper we consider only the quasicircular orbit (that is, circular apart from an adiabatic inspiral), and then the Kepler's third law is satisfied,
\begin{equation}
\omega=\left(\frac{\mathcal{G}m}{r^3}\right)^{1/2},
\end{equation}
where $\omega$ is the orbital frequency.

The gravitational radiation carries away the orbital energy of the binary system, which induces the increasing of the orbital frequency with time. Using the results of  \cite{Zhang2017},  the time derivate of the orbital frequency to leading order is given by,
\begin{equation}\label{wdot}
\dot{\omega}(t)=\frac{96}{5}(GM_c)^{\frac53}\omega^{\frac{11}{3}}\left[1+\frac{5}{192}(Gm\omega)^{-\frac23}\epsilon_d^2\right],
\end{equation}
where $\epsilon_d\equiv\epsilon_1-\epsilon_2$ is the difference in the screened parameter between the two objects. The first term in the square bracket is the contribution of the mass quadrupole radiation and the second term represents the scalar dipole radiation. When $\epsilon_d=0$, this result reduces to that of GR.

\section{Gravitational-wave waveforms in SMG}
\label{waveform}

In this section, for the general SMG,  we construct the time-domain GW waveforms, as well as their Fourier transforms using the stationary phase method.

\subsection{Time-domain waveforms}

In \cite{Zhang2017}, using the method of Green's function, the linearized field equations \eqref{linear_tensor_eq} and \eqref{linear_scalar_eq} have been solved in the wave zone. The metric perturbation is expressed in terms of the mass multipole moments and the scalar field is expressed in terms of the scalar multipole moments. Since we are calculating the lowest order waveform, analogous to GR, we need the metric perturbation only to quadrupole order. Similarly, for the scalar field, we need the scalar monopole, dipole and quadrupole moments.
The solutions of the tensor  and  scalar fields are given by \cite{Zhang2017}
\begin{equation}\label{h}
\bar{h}^{ij}=\frac{2G}{D}\frac{\partial^2}{\partial t^2}\sum_{a=1}^2 m_a r_a^i r^j_a \,,
\end{equation}
and
\begin{equation}\label{phi}
\varphi=-M_{\rm Pl}\frac{G}{D}\int_0^\infty dz J_1(z)\sum_{l=0}^2 \frac{1}{l !}N_L\partial_t^l {\mathcal M}_l^L \,,
\end{equation}
where $D$ is the coordinate distance from the compact binary to the observer, $J_1(z)$ is the Bessel function of the first kind and the capital letter $L$ is a multi-index and represents $l$ indices $i_1 i_2\cdots i_l$. The quantity $N_L$ is given by
\begin{equation}
N_L=N_{i_1}N_{i_2}\cdots N_{i_l} \,,
\end{equation}
where $N_i$ is the component of the direction unit vector $\hat{\mathbf{N}}$ of $D$.
The scalar multipole moments $\mathcal{M}_l^L$ are given by \footnote{Actually, the definition of $\mathcal{M}_l^L$  (Eq. (61) in \cite{Zhang2017}) includes contributions of the kinetic energy of the compact objects and gravitational binding energy between  them. But these corrections will not affect the GW waveforms to the required order in this paper, so we ignore these corrections. }
\begin{equation}
\mathcal{M}_l^L\equiv\mathcal{M}_l^{i_1 i_2 \cdots i_l}(t,D,z)
=\sum_a \epsilon_a \left[m_a r_a^L(t-D)-\frac{1}{u^{l+1}}m_a r_a^L(t-Du)\right]  \,,
\end{equation}
with $r_a^L(t)=r_a^{i_1}(t)r_a^{i_2}(t)\cdots r_a^{i_l}(t)$ and $u=\sqrt{1+\big(\frac{z}{m_s D}\big)^2}$. The calculations of the GW waveforms are based primarily on Eqs. \eqref{h} and \eqref{phi} which were obtained from the previous work \cite{Zhang2017}

Expressing the tensor  $\bar{h}^{ij}$  in terms of the relative displacement and velocity of the two compact objects of the binary system , we have
\begin{equation}
\bar{h}^{ij}=\frac{4G\mu}{D}\left[v^i v^j-\frac{\mathcal G m}{r^3} r^i r^j\right]_{t-D}.
\end{equation}
For the scalar field $\varphi$, retaining only terms to the order of $GM_{\rm Pl}mv^2/D$ in the monopole and quadrupole parts and to the order $GM_{\rm Pl}mv/D$ in the dipole term, we have
\begin{align}
\begin{split}\label{phi_wave}
\varphi(t,\mathbf{D})=&-\frac{G M_{\rm Pl} }{D}\int_0^\infty dz J_1(z)\left[\mathcal{M}_0+N_i\dot{\mathcal{M}}_1^i+\frac12 N_i N_j\ddot{\mathcal{M}}_2^{ij}\right]\\
=&-\frac{G M_{\rm Pl} }{D}\int_0^\infty dz J_1(z)\Bigg\{e^{-m_s D}(\epsilon_1m_1+\epsilon_2m_2)\\
&+\mu\epsilon_d\left[\hat{\mathbf{N}}\cdot\mathbf{v}(t-D)-\frac{1}{u^2}\hat{\mathbf{N}}\cdot\mathbf{v}(t-Du)\right]\\
&+\Gamma\left[-\frac{\mathcal{G}\mu m}{r^3}(\hat{\mathbf{N}}\cdot\mathbf{r})^2+\mu(\hat{\mathbf{N}}\cdot\mathbf{v})^2\right]_{t-D}\\
&-\frac{\Gamma}{u^3}\left[-\frac{\mathcal{G}\mu m}{r^3}(\hat{\mathbf{N}}\cdot\mathbf{r})^2+\mu(\hat{\mathbf{N}}\cdot\mathbf{v})^2\right]_{t-Du}\Bigg\},
\end{split}
\end{align}
where $\Gamma\equiv(\epsilon_1 m_2+\epsilon_2 m_1)/{m}$, $\mathbf{r}(t)=\mathbf{r_1}(t)-\mathbf{r_2}(t)$ is the relative coordinate and $\mathbf{v}(t)=\mathbf{v_1}(t)-\mathbf{v_2}(t)$ is the relative velocity of the two objects.
Note that the terms proportional to $\epsilon_d$ represent the scalar dipole contributions and the terms proportional to $\Gamma$ represent the scalar quadrupole contributions. Therefore, $\epsilon_d$ and $\Gamma$ are the indicators of the  scalar dipole and quadrupole moments, respectively. The monopole contribution takes the Yukawa form $e^{-m_s D}/D$ and is constant in time. Since we focus on the wavelike behavior of the scalar field in this article, the monopole contribution will be discarded in the following discussions. 

Comparing Eq. \eqref{phi_wave} with the scalar wave in the massless Brans-Dicke theory (Eqs. (5.2a) and (5.2b) in \cite{Lang-BD}), we find that when the compact binary system is in circular orbit, there are only three terms in the expression of the scalar wave in the massless Brans-Dicke theory. That is to say, the mass of the scalar field can double the number of terms in the scalar wave.
For the later convenience we express the scalar field as follows
\begin{equation}
\varphi(t,\mathbf{D})=\frac{\psi_1(t-D,\hat{\mathbf{N}})}{D}+\int_0^\infty dz J_1(z)\frac{1}{D}\left\{\frac{\psi_2(t-Du,\hat{\mathbf{N}})}{u^2}+\frac{\psi_3(t-Du,\hat{\mathbf{N}})}{u^3}\right\} \,,
\end{equation}
where
\begin{align}
\begin{split}
\psi_1(t-D,\hat{\mathbf{N}})&\equiv-G M_{\rm Pl}\left\{\epsilon_d\mu\hat{\mathbf{N}}\cdot\mathbf{v}+\Gamma\left[-\frac{\mathcal{G}\mu m}{r^3}(\hat{\mathbf{N}}\cdot\mathbf{r})^2+\mu(\hat{\mathbf{N}}\cdot\mathbf{v})^2\right]\right\}_{t-D},\\
\psi_2(t-Du,\hat{\mathbf{N}})&\equiv G M_{\rm Pl}\epsilon_d\mu\left(\hat{\mathbf{N}}\cdot\mathbf{v}\right)_{t-Du},\\
\psi_3(t-Du,\hat{\mathbf{N}})&\equiv G M_{\rm Pl}\Gamma\left[-\frac{\mathcal{G}\mu m}{r^3}\left(\hat{\mathbf{N}}\cdot\mathbf{r}\right)^2+\mu\left(\hat{\mathbf{N}}\cdot\mathbf{v}\right)^2\right]_{t-Du}.
\end{split}
\end{align}
Note that we have used the relation $\int_0^\infty dz J_1(z)=1$.

Under the influence of GWs \footnote{In this paper, we consider the  effects of GWs in the \textit{Jordan} frame. The overhead  bar  denotes the quantity in the \textit{Jordan} frame except the trace reversed metric perturbation.},
assuming that the distance between the test particles is less than the wavelength of the GWs and the test particles move slowly, we find that the separation of the test particles $\xi^i$ obeys the geodesic deviation equation
${d^2}\xi^i/{dt^2}=-\bar{R}_{0i0j}\xi^{j}$ \cite{maggiore2008gravitational}, where $\bar{R}_{0i0j}$ are the electric components of the Riemann tensor. Correspondingly, the GW field $\mathbf{h}_{ij}$ is defined by $\partial^2 \mathbf{h}_{ij}/{\partial t^2}=-2 \bar{R}_{0i0j}$ \cite{thorne1987gravitational}.

In a metric theory of gravity, there exist at most six polarization modes. When  a GW travels in the $\hat{\mathbf{N}}=\hat{z}$ direction, these polarizations can be expressed as
\begin{equation}
 \mathbf{h}_{ij}(t)=\left(
    \begin{matrix}
    {h}_b+{h}_+ & {h}_{\times} & {h}_x \\
    {h}_{\times} & {h}_b-{h}_+ & {h}_y \\
    {h}_x & {h}_{y} & {h}_L
    \end{matrix}
    \right).
\end{equation}
Note that the GW field $\mathbf{h}_{ij}$ differs from the metric perturbation $h_{ij}$ in general, although these two quantities can be derived from each other \cite{Yunes-BD}. Considering the displacement induced by the six polarizations on a sphere of test particles (see Figure 1 in \cite{PhysRevD.8.3308} or Figure 10.1 in \cite{will1993theory}), $h_+$, $h_\times$ and $h_b$ are purely transverse, $h_L$ is purely longitudinal, and $h_x$ and $h_y$ are mixed \cite{PhysRevD.8.3308}.
The response function $h(t)$ of a GW detector is a certain linear combination of the GW polarizations,
\begin{equation}
h(t)=\sum_A F_A h_A(t),
\end{equation}
where $A=+,\times,b,L,x,y,$ and $F_A$ is the detector antenna pattern function,
which depends on the geometry and orientation of the detector. Note that the  results in this paper can be applied to any antenna pattern function. In the next section we consider  Einstein Telescope (ET), a third-generation GW detector, as an example.
The detector antenna pattern functions of ET are given in Eqs. (C6)-(C13) in \cite{PhysRevD.95.124008}. 

We turn now to  the polarizations of GWs in SMG. Since the geodesic deviation equation only applies to the Jordan frame, we consider the Jordan frame metric
\begin{equation}\label{twoframe}
\bar{g}_{\mu\nu}=A^2(\phi)g_{\mu\nu}=A_0^2\left(\eta_{\mu\nu}+h_{\mu\nu}^{\rm TT}+\frac{2A_1}{A_0}\varphi \eta_{\mu\nu}\right).
\end{equation}
From the Jordan frame metric $\bar{g}_{\mu\nu}$, we can derive  the Jordan frame Riemann tensor $\bar{R}_{0i0j}$ straightforwardly,
\begin{equation}\label{Rie}
\bar{R}_{0i0j}=-\frac12A_0^2\left[-\frac{2A_1}{A_0}\varphi_{,ij}+\left(h_{ij}^{\rm TT}+\frac{2A_1}{A_0}\varphi \delta_{ij}\right)_{,00}\right]~.
\end{equation}
In order to obtain the GW polarizations from the Riemann tensor, we need to replace the spatial derivatives of the scalar field with the time derivative.
Using the relations
\begin{equation}
\partial_i\partial_j \left(\frac{\psi_1(t-D,\hat{\mathbf{N}})}{D}\right)=\frac{1}{D}N_i N_j \partial_t^2 \psi_1+\mathcal{O}\left(\frac{1}{D^2}\right),
\end{equation}
\begin{equation}
\partial_i\partial_j \left(\frac{\psi_2(t-P,\hat{\mathbf{N}})}{Du^2}\right)=\frac{1}{Du^2}N_iN_j\left(\frac{d P}{dD}\right)^2 \partial_t^2 \psi_2+\mathcal{O}\left(\frac{1}{D^2}\right),
\end{equation}
\begin{equation}
\partial_i\partial_j \left(\frac{\psi_3(t-P,\hat{\mathbf{N}})}{Du^3}\right)=\frac{1}{Du^3}N_iN_j\left(\frac{d P}{dD}\right)^2 \partial_t^2 \psi_3+\mathcal{O}\left(\frac{1}{D^2}\right),
\end{equation}
with $P\equiv Du$ and $dP/d D=1/u$, we have
\begin{align}
\begin{split}
\bar{R}_{0i0j}=&-\frac12 A_0^2 \frac{\partial^2}{\partial t^2}\Bigg\{h_{ij}^{TT}+(\delta_{ij}-N_i N_j)\frac{2A_1}{A_0}\varphi
-N_i N_j\frac{2A_1}{A_0}\frac{1}{D}\int_0^\infty dz J_1(z)\left(\frac{1}{u^2}-1\right)\left(\frac{\psi_2}{u^2}+\frac{\psi_3}{u^3}\right)\Bigg\}.
\end{split}
\end{align}
The factor $A_0^2$ in Eq. \eqref{twoframe} should be absorbed by a coordinate rescaling
$x'^\mu=A_0x^\mu$. In the $x'^\mu$ coordinates, the Jordan frame Riemann tensor is
\begin{align}\label{rie_gw}
\begin{split}
\bar{R}'_{0i0j}=&-\frac12 \frac{\partial^2}{\partial t'^2}\Bigg\{h_{ij}^{TT}+(\delta_{ij}-N_i N_j)\frac{2A_1}{A_0}\varphi
-N_i N_j\frac{2A_1}{A_0}\frac{1}{D}\int_0^\infty dz J_1(z)\left(\frac{1}{u^2}-1\right)\left(\frac{\psi_2}{u^2}+\frac{\psi_3}{u^3}\right)\Bigg\}\\
=&-\frac12 \frac{\partial^2}{\partial t'^2}\Big\{(\delta_{ij}-N_i N_j)h_b+N_iN_j h_L+h_{ij}^{TT}\Big\}.
\end{split}
\end{align}
 We observe that the massive scalar field induces two polarizations, $h_b$ and $h_L$. Due to the existence of the longitudinal polarization $h_L$, $\Psi_2$ component of the Weyl tensor is nonzero and SMG is of class $II_6$ in  the $E(2)$ classification \cite{will1993theory,PhysRevD.8.3308,PhysRevLett.30.884}. In SMG, there are three dynamical degrees of freedom (i.e., two tensor degrees and one scalar degree), but four GW polarization modes. This is an excellent illustration of a discrepancy between the number of  polarizations in the $E(2)$ classification and the number of dynamical degrees of freedom.

From the Riemann tensor \eqref{rie_gw}, we can identify the waveforms of the four polarizations of GWs in SMG.
The breathing polarization is \footnote{The breathing polarization is also called conformal polarization \cite{deRham2014}.}
\begin{align}\label{hb}
\begin{split}
h_b(t)=&-\frac{2A_1}{A_0}\frac{GM_{\rm Pl}}{D} \Big\{\mu \epsilon_d v \sin \theta \cos(\Phi)
+ \Gamma \mathcal{G}^{2/3}M_c^{5/3}\omega^{2/3}\sin^2\theta \cos(2\Phi) \Big\}_{t-D}\\
&+\frac{2A_1}{A_0}\frac{G M_{\rm Pl}}{D}\int_0^\infty dz J_1(z)\left\{\frac{1}{u^2}\mu\epsilon_d v\sin \theta \cos(\Phi)
+\frac{1}{u^2}\Gamma\mathcal{G}^{2/3}M_c^{5/3}\omega^{2/3}\sin^2\theta\cos(2\Phi)\right\}_{t-Du},
\end{split}
\end{align}
and the longitudinal polarization is
\begin{align}\label{hl}
h_L(t)=-\frac{2A_1}{A_0}\frac{G M_{\rm Pl}}{D}\int_0^\infty dz J_1(z)\left(\frac{1}{u^2}-1\right)\Bigg[\frac{1}{u^2}\mu\epsilon_d v\sin \theta \cos(\Phi)
+\frac{1}{u^2}\Gamma\mathcal{G}^{2/3}M_c^{5/3}\omega^{2/3}\sin^2\theta\cos(2\Phi)\Bigg]_{t-Du}.
\end{align}
The waveforms of the tensor polarizations are
\begin{equation}\label{h+}
h_+(t)=-\left(1+\frac12\epsilon_1\epsilon_2\right)^{\frac23}~~\frac{4(GM_c)^{5/3}\omega^{2/3}}{D}~~\frac{1+\cos^2\theta}{2}\cos(2\Phi)\Big|_{t-D},
\end{equation}
\begin{equation}\label{hx}
h_\times(t)=-\left(1+\frac12\epsilon_1\epsilon_2\right)^{\frac23}~~\frac{4(GM_c)^{5/3}\omega^{2/3}}{D}~~\cos\theta\sin(2\Phi)\Big|_{t-D},
\end{equation}
where $\Phi(t)=\int^t \omega(t')dt'$  is the orbital phase of the binary system, $\theta$  is  the inclination angle of the binary orbital angular momentum along the line of sight,
$M_c \left(\equiv \mu^{3/5}m^{2/5}\right)$ is the chirp mass. Note that we have used the relations $\hat{\mathbf{N}}\cdot\mathbf{v}=v\sin \theta \cos\Phi$, $\hat{\mathbf{N}}\cdot\mathbf{r}=r\sin \theta \sin\Phi$ and $v=\omega r$.

We perform the integrals containing the Bessel function in $h_b$ and $h_L$ in the limit $D\rightarrow
\infty$. The detailed steps are discussed in Appendix \ref{appendix_a}. After performing these integrals, we derive the waveform of the breathing polarization,
\begin{equation}
h_b=h_{b1}+h_{b2},
\end{equation}
\begin{eqnarray}
h_{b1}(t)&=&-\frac{2A_1}{A_0}\frac{G M_{\rm Pl}}{D}\mu\epsilon_d(\mathcal{G}m\omega)^\frac13 v_s(\omega)\sin\theta
\cos\left(\frac{m_s^2 D}{\sqrt{\omega^2-m_s^2}}+\Phi\right)\Bigg|_{t-Du_1}~,
\label{hb1} \\
h_{b2}(t)&=&-\frac{2A_1}{A_0}\frac{G M_{\rm Pl}}{D} \Gamma\mathcal{G}^{2/3}M_c^{5/3}\omega^{2/3}v_s(2\omega)^2\sin^2\theta
\cos\left(\frac{m_s^2 D}{\sqrt{4\omega^2-m_s^2}}+2\Phi\right)\Bigg|_{t-Du_2}~, \label{hb2}
\end{eqnarray}
and the waveform of the  longitudinal polarization,
\begin{equation}
h_L=h_{L1}+h_{L2},
\end{equation}
\begin{eqnarray}
h_{L1}(t)&=&-\frac{m_s^2}{\omega^2}\frac{2A_1}{A_0}\frac{G M_{\rm Pl}}{D}\mu\epsilon_d(\mathcal{G}m\omega)^\frac13 v_s(\omega)\sin\theta
\cos\left(\frac{m_s^2 D}{\sqrt{\omega^2-m_s^2}}+\Phi\right)\Bigg|_{t-Du_1}~, \label{hl1} \\
h_{L2}(t)&=&-\frac{m_s^2}{4\omega^2}\frac{2A_1}{A_0}\frac{G M_{\rm Pl}}{D} \Gamma\mathcal{G}^{2/3}M_c^{5/3}\omega^{2/3}v_s(2\omega)^2\sin^2\theta\cos\left(\frac{m_s^2 D}{\sqrt{4\omega^2-m_s^2}}+2\Phi\right)\Bigg|_{t-Du_2}~, \label{hl2}
\end{eqnarray}
where $u_n=\left.{n\omega}/{\sqrt{n^2\omega^2-m_s^2}}\right|_{t-D}$ and $v_s(\omega)=\sqrt{1-{m_s^2}/{\omega^2}}$ is the speed of the scalar wave with frequency $\omega$, which is smaller than the speed of light \footnote{To avoid  the severe constraints from  the vacuum gravi-\v{C}erenkov radiation by matter such as cosmic rays \cite{Elliott2006}, one normally
requires ${m_s^2}/{\omega^2}\ll 1$.}.

We find that, to the required order, both the breathing polarization $h_b$ and the longitudinal polarization $h_L$ have two frequency modes. In addition, the amplitude of $h_L$ decreases with time, while other polarizations all chirp (that is, both of their amplitudes and frequencies increase with time). Since $h_{b1}$ and $h_{L1}$ are proportional to $\epsilon_d$, they stem from the scalar dipole radiation as mentioned above. Similarly, since $h_{b2}$ and $h_{L2}$ are proportional to $\Gamma$, they stem from the scalar quadrupole radiation. In particular, we find the simple linear relationships between $h_b$ and $h_L$, given  by,
\begin{equation}\label{linear}
h_{L1}=\frac{m_s^2}{\omega^2}h_{b1}, \quad h_{L2}=\frac{m_s^2}{4\omega^2}h_{b2}.
\end{equation}
These relations are the direct consequence of the linearized field equation \eqref{linear_scalar_eq}, which can be understood as follows: Considering a wave packet $\varphi(t,D,\hat{\mathbf{N}})$ centered at a frequency $\omega_{\rm GW}$, from Eq. \eqref{linear_scalar_eq} we find
\begin{equation}\label{mw}
\partial_D^2 \varphi=\left(1-\frac{m_s^2}{\omega_{\rm GW}^2}\right)\partial_t^2\varphi,
\end{equation}
where we have used the relations $\partial_i\varphi=N_i\partial_D\varphi$ and $\partial_t^2\varphi=-\omega_{\rm GW}^2\varphi$. Applying Eq. \eqref{mw} to the electric components of the Riemann tensor \eqref{Rie}, we have
\begin{equation}
\bar{R}_{0i0j}=-\frac12A_0^2\left[-\frac{2A_1}{A_0}N_i N_j\left(1-\frac{m_s^2}{\omega_{\rm GW}^2}\right)\varphi_{,00}+\left(h_{ij}^{\rm TT}+\frac{2A_1}{A_0}\varphi \delta_{ij}\right)_{,00}\right]~.
\end{equation}
Consequently,
\begin{equation}
h_b=\frac{2A_1}{A_0}\varphi,\quad h_L=\frac{m_s^2}{\omega_{\rm GW}^2}\frac{2A_1}{A_0}\varphi=\frac{m_s^2}{\omega_{\rm GW}^2}h_b~.
\end{equation}
{This linear relation has also been obtained in the case of plane waves in} \cite{PhysRevD.62.024004}.
If $\omega_{\rm GW}$ is in the bandwidth of the ground-based detectors, $\omega_{\rm GW}\simeq 100{\rm Hz}$, and the reduced Compton wavelength of the scalar field is roughly of the cosmological scales, $m_s^{-1}\simeq {\rm 1Mpc}$, then $m_s^2/\omega_{\rm GW}^2\simeq 10^{-32}$. Therefore, it is very hard to detect the longitudinal polarization $h_L$.

Having obtained the amplitude ratio between the two scalar polarizations, we now turn to discuss the amplitude ratio between the scalar polarizations and the tensor polarizations. It follows immediately from Eqs. \eqref{h+}, \eqref{hb1} and \eqref{hb2} that the amplitude ratios of $h_{b1}$ to $h_+$ and $h_{b2}$ to $h_+$ are
\begin{equation}
\frac{|h_{b1}|}{|h_+|}\approx\frac{A_1 M_{\rm Pl}}{A_0}\times\frac{\epsilon_d}{v}\times\frac{\sin \theta}{1+\cos^2\theta},\quad \frac{|h_{b2}|}{|h_+|}\approx\frac{A_1 M_{\rm Pl}}{A_0}\times\Gamma\times\frac{\sin^2 \theta}{1+\cos^2\theta}~.
\end{equation}
When the GW emitted by the compact binary enters the bandwidth of the  ground-based detector, the relative velocity of the compact binary $v$ is of order 0.1. As a result, the relative intensity of $h_{b1}$ and $h_{b2}$ is controlled by $\epsilon_d$ and $\Gamma$. For the binary neutron star (BNS) system or binary white dwarf (BWD) system, if we assume that the screened parameters of NSs or WDs are the same, then $\epsilon_d\sim0$ and $h_{b2}$ is dominant over $h_{b1}$, that is, the quadrupole contribution is dominant over the dipole contribution in this situation. For the binary black hole (BBH) systems, since the sensitivity of  BH is zero (see Appendix \ref{sbh}) and $\epsilon_d=\Gamma=0$, there is no scalar radiation. Meanwhile, the tensor polarizations also reduce to those of GR. Sotiriou and Faraoni \cite{PhysRevLett.108.081103} have proved that isolated BHs in scalar-tensor gravity are not different from those given in GR. Our results suggest that, up to the quadrupole order, the inspiral BBH systems in scalar-tensor gravity are also the same as those in GR. On the other hand, for the NS-BH binaries, since $\epsilon_{\rm BH}=0$, $\epsilon_d$ and $\Gamma$ are in the same order of the magnitude, we find that $|h_{b1}|$ is  several times larger than $|h_{b2}|$.  Similar results also apply to the WD-BH  and NS-WD systems.

\subsection{Waveforms in the stationary phase approximation}

In GW data analysis, one often works with the Fourier transforms of the GW waveforms. During the inspiral, the change in orbital frequency over a single period is negligible, and we can apply the stationary phase approximation (SPA) to compute the Fourier transform. Now we take the plus polarization $h_+$ as an example to illustrate SPA. The Fourier transform of $h_+(t)$ is
\begin{equation}
\tilde{h}_+(f)=\int h_+(t'/A_0)e^{i 2\pi f t'} dt'=A_0\int h_+(t)e^{i 2\pi fA_0 t} dt,
\end{equation}
where $A_0$ comes from the coordinate rescaling.
Substitution of  Eq. \eqref{h+} into the above equation yields
\begin{align}
\begin{split}
\tilde{h}_+(f)=&-A_0(1+\frac12\epsilon_1\epsilon_2)^{\frac23}\times\frac{4(GM_c)^{5/3}}{D}\times\frac{1+\cos^2\theta}{2}\\
&\times\frac12 e^{i2\pi fA_0 D}\int \omega(t)^{2/3}\left[e^{i(-2\Phi(t)+2\pi fA_0t)}+e^{i(2\Phi(t)+2\pi fA_0t)}\right] dt~.
\end{split}
\end{align}
The second term in the square bracket does not have a stationary point, i.e., a value of $t$ satisfying $d(2\Phi(t)+2\pi fA_0t)/dt=0$. Thus, the second term is always oscillating fast and its integration can be neglected.

The stationary phase point of the first term $t_*$ is determined by
\begin{equation}
\frac{d}{dt}(-2\Phi+2\pi f A_0 t)\Big|_{t=t_*}=0,\quad \omega(t_*)=\pi f A_0~.
\end{equation}
Expanding the exponential around $t_*$ to second order,
\begin{equation}
-2\Phi(t)+2\pi fA_0 t=-2\Phi(t_*)+2\pi fA_0t_*-\dot{\omega}(t_*)(t-t_*)^2+\cdots,
\end{equation}
we obtain $\tilde{h}_+(f)$ analytically
\begin{align}
\tilde{h}_+(f)=-A_0\left(1+\frac12\epsilon_1\epsilon_2\right)^{\frac23}~\frac{4(GM_c)^{5/3}}{D}~\frac{1+\cos^2\theta}{2}
\times\frac12 \omega(t_*)^{2/3}\sqrt{\frac{\pi}{\dot{\omega}(t_*)}}e^{i\Psi_+},
\end{align}
with the phase $\Psi_+=2\pi fA_0(D+t_*)-2\Phi(t_*)-{\pi}/{4}$.

Using the time derivative of the orbital frequency in Eq. \eqref{wdot}, we can eliminate $t_*$ in the phase $\Psi_+$ in terms of the frequency $f$,
\begin{align}
\begin{split}
&2\pi fA_0 t_*-2\Phi(t_*)\\
=&\int_{t_c}^{t_*}\left(2\pi fA_0-2\omega(t)\right)dt +2\pi fA_0t_c-2\Phi_c\\
=&\int_{\omega(t_c)}^{\omega(t_*)}\left(2\pi fA_0-2\omega\right)\frac{d\omega}{\dot{\omega}}+2\pi fA_0t_c-2\Phi_c\\
=&\frac{3}{128}(GM_c\pi f A_0)^{-\frac53}\left[1-\frac{5}{336}(Gm\pi fA_0)^{-\frac23}\epsilon_d^2\right]+2\pi fA_0t_c-2\Phi_c,
\end{split}
\end{align}
where $t_c$ is the time at which $\omega\to\infty$ and $\Phi_c=\Phi(t_c)$.

Combining the above results, we find the Fourier transform of the plus polarization,
\begin{align}\label{hplus}
\begin{split}
\tilde{h}_+(f)=&-\left(1+\frac12\epsilon_1\epsilon_2\right)^{\frac23}\left(\frac{5\pi}{24}\right)^{\frac12}\times\frac{A_0(GM_c)^{5/6}}{D}\times\frac{1+\cos^2 \theta}{2}(\pi fA_0)^{-7/6}\\
&\times\left[1-\frac{5}{384}(Gm\pi fA_0)^{-2/3}\epsilon_d^2\right]e^{i\Psi_+},
\end{split}
\end{align}
with the phase $\Psi_+=2\pi fA_0(D+t_c)-2\Phi_c-\frac{\pi}{4}+\frac{3}{128}\left(GM_c\pi f A_0\right)^{-\frac53}\left[1-\frac{5}{336}(Gm\pi fA_0)^{-\frac23}\epsilon_d^2\right]$.
When $\epsilon_1=\epsilon_2=0$ and $A_0=1$, the expression of $\tilde{h}_+(f)$ reduces to that of GR.

Following a similar procedure, we can derive the Fourier transforms of other polarizations. In particular,
the cross polarization is
\begin{align}
\begin{split}
\tilde{h}_\times(f)=&-(1+\frac12\epsilon_1\epsilon_2)^{\frac23}\Big(\frac{5\pi}{24}\Big)^{\frac12}\times\frac{A_0(GM_c)^{5/6}}{D}\times\cos\theta(\pi fA_0)^{-7/6}\\
&\times\left[1-\frac{5}{384}(Gm\pi fA_0)^{-2/3}\epsilon_d^2\right]e^{i\Psi_\times},
\end{split}
\end{align}
with the phase $\Psi_\times=\Psi_++{\pi}/{2}$.

The Fourier transform of the breathing polarization is
\begin{equation}
\tilde{h}_b(f)=\tilde{h}_{b1}(f)+\tilde{h}_{b2}(f),
\end{equation}
where
\begin{align}
\begin{split}
\tilde{h}_{b1}(f)=&-\frac{5\pi}{48}A_1 M_p\frac{G\mu}{D}\epsilon_d(\mathcal{G}m)^{\frac13}(GM_c)^{-\frac{5}{6}}(2\pi fA_0)^{-\frac32}\sin\theta\\
&\times\left[1-32m_s^2 D(GM_c)^{\frac53}(2\pi fA_0)^{\frac23}-\frac{5}{384}(Gm2\pi fA_0)^{-\frac23}\epsilon_d^2-\frac{m_s^2}{2(2\pi fA_0)^2}\right]e^{i\Psi_{b1}},
\end{split}
\end{align}
\begin{align}
\begin{split}
\tilde{h}_{b2}(f)=&-\frac12\Big(\frac{5\pi}{24}\Big)^{\frac12}A_1M_p\frac{GM_c}{D}\Gamma(\mathcal{G}M_c)^{\frac23}(GM_c)^{-\frac56}(\pi fA_0)^{-\frac{7}{6}}\sin^2 \theta\\
&\times\left[1-\frac{22}{5}(GM_c)^{\frac53}m_s^2D(\pi fA_0)^{\frac23}-\frac{5}{384}(Gm\pi fA_0)^{-\frac23}\epsilon_d^2-\frac{m_s^2}{4(\pi fA_0)^2}\right]e^{i\Psi_{b2}},
\end{split}
\end{align}
with the corresponding phases 
\begin{align}
\begin{split}
\Psi_{b1}(f)=&2\pi fA_0(D+t_c)-\frac{m_s^2 D}{4\pi fA_0}-\frac{\pi}{4}-\Phi_c\\
&+\frac{3}{256}(2\pi fA_0 GM_c)^{-\frac53}\Bigg[1-\frac{5}{336}\eta^{\frac25}\epsilon_d^2(GM_c2\pi fA_0)^{-\frac23}\Bigg],
\end{split}
\end{align}
\begin{align}
\begin{split}
\Psi_{b2}(f)=&2\pi fA_0(D+t_c)-\frac{m_s^2 D}{4\pi fA_0}-\frac{\pi}{4}-2\Phi_c\\
&+\frac{3}{128}(\pi fA_0 GM_c)^{-\frac53}\Bigg[1-\frac{5}{336}\eta^{\frac25}\epsilon_d^2(GM_c\pi fA_0)^{-\frac23}\Bigg]\\
=&\Psi_+-\frac{m_s^2R}{4\pi fA_0}.
\end{split}
\end{align}
$\eta=\mu/m$ is the symmetric mass ratio.
The Fourier transform of the longitudinal polarization is
\begin{equation}
\tilde{h}_L(f)=\tilde{h}_{L1}(f)+\tilde{h}_{L2}(f),
\end{equation}
where
\begin{align}
\begin{split}
\tilde{h}_{L1}(f)=&-\left(\frac{5\pi}{48}\right)^{\frac12}A_1 M_p\frac{G\mu}{D}\epsilon_d(\mathcal{G}m)^{\frac13}m_s^2(GM_c)^{-\frac56}(2\pi f A_0)^{-\frac72}\sin\theta\\
&\times\Bigg[1-\frac{256}{5}(GM_c)^{\frac53}m_s^2D(2\pi fA_0)^{\frac23}-\frac{5}{384}(Gm2\pi fA_0)^{-\frac23}\epsilon_d^2-\frac{m_s^2}{2(2\pi fA_0)^2}\Bigg]e^{i\Psi_{L1}},
\end{split}
\end{align}
\begin{align}
\begin{split}
\tilde{h}_{L2}(f)=&-\frac18\Big(\frac{5\pi}{24}\Big)^{\frac12}A_1M_p\frac{GM_c}{D}\Gamma(\mathcal{G}M_c)^{\frac23}m_s^2(GM_c)^{-\frac56}(\pi fA_0)^{-\frac{19}{6}}\sin^2 \theta\\
&\times\Bigg[1-\frac{22}{5}(GM_c)^{\frac53}m_s^2D(\pi fA_0)^{\frac23}-\frac{5}{384}(Gm\pi fA_0)^{-\frac23}\epsilon_d^2-\frac{m_s^2}{4(\pi fA_0)^2}\Bigg]e^{i\Psi_{L2}},
\end{split}
\end{align}
with the corresponding phases
\begin{equation}
\Psi_{L1}=\Psi_{b1},
\end{equation}
\begin{equation}\label{psil2}
\Psi_{L2}=\Psi_{b2}=\Psi_+-\frac{m_s^2D}{4\pi fA_0}.
\end{equation}
Note that $\tilde{h}_L(f)$ has the same phases as $\tilde{h}_b(f)$ because of  the linear relations \eqref{linear}. The phase difference  $-\frac{m_s^2D}{4\pi fA_0}$, which takes the form predicted by Will \cite{PhysRevD.57.2061}, is a result of the mass of the scalar field.

The response function of GW detectors in SMG is given by
\begin{equation}
h(t)=F_\times h_\times(t)+F_+ h_+(t)+F_bh_b(t)+F_L h_L(t),
\end{equation}
and the corresponding Fourier transform is
\begin{align}
\begin{split}
\tilde{h}(f)=&F_\times \tilde{h}_\times(f)+F_+\tilde{h}_+(f)+F_b\tilde{h}_b(f)+F_L\tilde{h}_L(f)\\
\equiv&\tilde{h}^{(1)}(f)+\tilde{h}^{(2)}(f),
\end{split}
\end{align}
where $\tilde{h}^{(1)}(f)=F_b \tilde{h}_{b1}(f)+F_L\tilde{h}_{L1}(f)$ and $\tilde{h}^{(2)}(f)=F_\times\tilde{h}_\times(f)+F_+\tilde{h}_+(f)+F_b\tilde{h}_{b2}(f)+F_L\tilde{h}_{L2}(f)$.

Note that in Eqs. \eqref{hplus}-\eqref{psil2} the distance $D$, the masses $m$, $\mu$, $M_c$, $m_s$ and the time $t_c$ are in the Einstein frame, which can be transformed into the Jordan frame by the relations \cite{Zhang2016},
\begin{equation}
\bar{t}_c=A_0 t_c,~~\bar{D}=A_0 D,~~\bar{m}=m/A_0,~~\bar{\mu}=\mu/A_0,~~\bar{M}_c=M_c/A_0.
\end{equation}
Combining  Eqs. \eqref{hplus}-\eqref{psil2} and using the above relations, we obtain
\begin{align}\label{h1}
\begin{split}
\tilde{h}^{(1)}(f)=&\frac{(G\bar{M}_c)^{\frac56}}{\bar{D}}\left(\frac{5}{48}\right)^{\frac12}\pi^{-\frac12}(2f)^{-\frac76}\Bigg[-\frac{J}{2}(G\bar{m}\bar{m}_s)^2(2\pi fG\bar{m})^{-\frac{13}{3}}\\
&-\frac{5}{384}JA_0^{-\frac43}\epsilon_d^2(2\pi fG\bar{m})^{-3}+\left(J-\frac{E}{2}(G\bar{m}\bar{m}_s)^2\right)(2\pi fG\bar{m})^{-\frac73}\\
&-\frac{256}{5}J \bar{m}_s^2\bar{D}G\bar{m}\eta A_0^{\frac{10}{3}}(2\pi fG\bar{m})^{-\frac53}-\frac{5}{384}EA_0^{-\frac43}\epsilon_d^2(2\pi fG\bar{m})^{-1}\\
&+E(2\pi fG\bar{m})^{-\frac13}-32E\bar{m}_s^2\bar{D}G\bar{m}\eta A_0^{\frac{10}{3}}(2\pi fG\bar{m})^{\frac13}\Bigg]\\
&\times \exp\left\{{i\left[2\pi f(\bar{D}+\bar{t}_c)-\frac{\pi}{4}+\psi(f)-\frac{\bar{m}_s^2\bar{D}}{4\pi f}\right]}\right\},
\end{split}
\end{align}

\begin{align}\label{h2}
\begin{split}
\tilde{h}^{(2)}(f)=&\left(\frac{5}{96}\right)^{\frac12}\pi^{-\frac23}\frac{(G\bar{M}_c)^{\frac56}}{\bar{D}}f^{-\frac76}\Bigg\{T\Bigg[-\frac{F_L}{16}(G\bar{m}\bar{m}_s)^4(\pi fG\bar{m})^{-4}\\
&+\frac{F_L}{4}(G\bar{m}\bar{m}_s)^2S_{-1}(\pi fG\bar{m})^{-\frac83}\\
&+\frac14(G\bar{m}\bar{m}_s)^2(F_L-F_b)(\pi fG\bar{m})^{-2}\\
&-\frac{11}{10}F_L(G\bar{m}\bar{m}_s)^3\bar{m}_s \bar{D}\eta A_0^{\frac{10}{3}}(\pi fG\bar{m})^{-\frac43}\\
&+F_b S_{-1}(\pi fG\bar{m})^{-\frac23}+F_b\\
&-\frac{22}{5}F_b G\bar{m}\bar{m}_s^2\bar{D}\eta A_0^{\frac{10}{3}}(\pi fG\bar{m})^{\frac23}\Bigg]\\&+\Bigg[Q+QS_{-1}(G\bar{m}\pi f)^{-\frac23}\Bigg]e^{-i\varphi_{(2,0)}}P_{(2,0)}\Bigg\}\\
&\times  \exp\left\{{i[2\pi f(\bar{D}+\bar{t}_c)-\frac{\pi}{4}+2\psi(f/2)]}\right\},
\end{split}
\end{align}
where
\bqn
E&=&-F_bA_1M_{\rm Pl} \epsilon_d\sin\theta\left(1+\frac12\epsilon_1\epsilon_2\right)^{1/3},\nb\\
J&=& -F_L A_1 M_{\rm Pl} \epsilon_d\sin\theta(G\bar{m}\bar{m}_s)^2\left(1+\frac12\epsilon_1\epsilon_2\right)^{\frac13},\nb\\
Q&=& A_0^{5/3}\left(1+\frac12\epsilon_1\epsilon_2\right)^{2/3},\quad S_{-1}= -\frac{5}{384}\epsilon_d^2A_0^{-4/3},\nb\\
T&=& -A_0^{2/3}A_1 M_{\rm Pl} \Gamma\left(1+\frac12\epsilon_1\epsilon_2\right)^{2/3}\sin^2\theta e^{-i\frac{\bar{m}_s^2\bar{D}}{4\pi f}},\nb\\
\psi(f)&=& \frac{3}{256}(2\pi fA_0^2 G\bar{M}_c)^{-\frac53}\left[1-\frac{5}{336}\epsilon_d^2 A_0^{-\frac43}(G\bar{m}2\pi f)^{-\frac23}\right]-\Phi_c.
\eqn
Similar to \cite{PhysRevD.95.124008}, we have defined $e^{-i\varphi_{(2,0)}}P_{(2,0)}\equiv -[F_+(1+\cos^2\theta)+2iF_\times\cos\theta]$.

Considering the results of the Solar System experiments, we have constrained  $|A_0-1|$ to be less than $10^{-10}$ in the Milky Way background in various specific models of SMG \cite{Zhang2016}.
So, it is natural to assume that $A_0$ cannot deviate from unity too much in the background of other galaxies, e.g. the host galaxy for a GW event. Therefore, we will set $A_0=1$ in the following
discussion and the overhead bars in Eqs. \eqref{h1} and \eqref{h2} can be dropped.

\section{Parametrized post-Einsteinian parameters}
\label{ppe}

In the standard ppE framework, one considers possible deviation of the two tensor polarizations $(h_+,h_\times)$ from the GR predictions.
As Yunes and Pretorius found in \cite{PhysRevD.80.122003}, the Fourier transform of the response function in metric theories of gravity can be generically cast in the form,
\begin{equation}
\tilde{h}(f)=\tilde{h}_{\rm GR} (f)\left(1+\alpha(\pi M_c f)^\frac{a}{3}\right)e^{i \beta(\pi M_c f)^\frac{b}{3}},
\end{equation}
where $(\alpha,\beta,a,b)$ are the four ppE parameters and $\tilde{h}_{\rm GR} (f)$ denotes the GR prediction of the Fourier transform of the response function. $\alpha(\pi M_c f)^\frac{a}{3}$ denotes the non-GR correction to the GW amplitude
while $\beta(\pi M_c f)^\frac{b}{3}$ corresponds to that to the GW phase \cite{tahura2018parameterized}. For instance,
the ppE parameters of Brans-Dicke theory are $(\alpha_{\rm BD}, \beta_{\rm BD}, a_{\rm BD}, b_{\rm BD})=(\frac{112}{3}\beta_{\rm BD},-\frac{5}{3584}\eta^{\frac25}(s_1-s_2)^2\frac{1}{2+\omega_{\rm BD}},-2,-7)$,
where $s_1,s_2$ are the sensitivities of the compact objects in Brans-Dicke theory and $\omega_{\rm BD}$ is the coupling constant \cite{Yunes-BD}.

Since the standard ppE framework only includes the  two tensor polarizations $(h_+,h_\times)$,
to obtain the ppE parameters in SMG, we focus on  the two tensor polarizations $(h_+,h_\times)$, and the Fourier transform of the response function becomes
\begin{align}\label{fh}
\begin{split}
F_+\tilde{h}_++F_\times \tilde{h}_\times=\left(\frac{5}{96}\right)^{1/2} \pi^{-2/3}\frac{(GM_c)^{5/6}}{D} f^{-7/6}\left[Q+QS_{-1}(Gm\pi f)^{-2/3}\right]e^{-\Psi_+}e^{-i \varphi_{(2,0)} }P_{(2,0)},
\end{split}
\end{align}
where $Q=(1+\frac12\epsilon_1\epsilon_2)^{2/3}$, $S_{-1}=-\frac{5}{384}\epsilon_d^2$
\noindent and $\Psi_+=2\pi f(D+t_c)-\frac{\pi}{4}+\frac{3}{128}(\pi f GM_c)^{-5/3}\left[1-\frac{5}{336}\epsilon_d^2(Gm\pi f)^{-2/3}\right]-2\Phi_c$. From the formula, we can identify the ppE parameters in SMG as follows,
\begin{align}\label{ppe_smg}
\alpha=-\frac{5}{384}\epsilon_d^2\eta^{2/5}, \quad \beta=-\frac{5}{14336}\epsilon_d^2\eta^{2/5}, \quad
a=-2, \quad b=-7~,
\end{align}
where the coefficient $Q$ has been absorbed into the definition of $G$.
The same as that of Brans-Dicke theory, we obtain the ratio  ${\alpha}/{\beta}={112}/{3}$,
which is a result of  the fact that the non-GR corrections to the Fourier transform of the tensor polarizations in these two theories all originate from the dipole radiation in the GW energy flux  \cite{Yunes-BD}. The ppE parameters $\alpha$ and $\beta$ in these two theories all depend on the difference between the scalar charges and the symmetric mass ratio $\eta$. In the test mass limit ($\eta\to0$), $\alpha$ and $\beta$  become zero in these two theories. Since the extend ppE framework does not have enough parameters to parametrize Eqs. \eqref{h1} and \eqref{h2} \cite{Yunes-BD} and the tensor polarizations are dominant over the scalar polarizations, we will not apply the extended ppE framework to SMG.

It is important to emphasize that, the results derived above are quite general, which are applicable for any SMG model and for any kind of compact binary systems. Therefore, we expect the observations of gravitational radiation by various compact binaries, in particular the asymmetric binaries, could place constraints on the SMG theories. For instance, the future space-based LISA mission could detect the GW signals of NS-BH binaries, WD-BH binaries, BH-main sequence (BH-MS) binaries, as well as NS-WD binaries, which provide the excellent opportunity to constrain the sensitivities of NSs, WDs and MSs. In this paper, we consider only the GW signals from the inspiral NS-BH binaries, observed by the ground-based ET, to constrain the SMG theories, and leave the other potential constraints as a future work.

In previous work \cite{PhysRevD.95.124008}, we found that, by observing the GWs of NS-BH binaries up to redshift $z=5$, ET could potentially place the stringent constraints on the Brans-Dicke theory, and the bound on the coupling constant $\omega_{\rm BD}$ could be $\omega_{\rm BD}>10^6\times({N_{\rm GW}}/{10^4})^{1/2}$,
where $N_{\rm GW}$ is the total number of observed GW events, and the sensitivities of the compact objects are fixed to be $s_1=0.5$ for BH and $s_2=0.2$ for NS. As illustrated in \cite{PhysRevD.95.124008}, this constraint is dominant by the non-GR contribution of GW phases through ppE parameter $\beta_{\rm BD}$. So, the bound on $\omega_{\rm BD}$  can be translated into a constraint on  $\beta_{\rm BD}$ as follows,
\begin{equation}
|\beta_{\rm BD}|<1.3\times10^{-10}\eta^{2/5}\left(\frac{10^4}{N_{\rm GW}}\right)^{\frac12}.
\end{equation}
Since the ppE parameters in SMG are quite similar to those in Brans-Dicke theory, in particular the values of $a$ and $b$ are exactly the same for both theories, we anticipate that ET could also place  constraints on the ppE parameter $\beta$ of SMG at the same level,
\begin{equation}\label{ppe_beta}
|\beta|=\frac{5}{14336}\epsilon_d^2\eta^{2/5}<1.3\times10^{-10}\eta^{2/5}\left(\frac{10^4}{N_{\rm GW}}\right)^{\frac12},
\end{equation}
that is
$|\epsilon_d|<6\times10^{-4}\times({10^4}/{N_{\rm GW}})^{1/4}$ for NS-BH binary system.

The scalar field outside a single BH in SMG is \cite{Zhang2017}
\begin{equation}
\phi=\phi_0+\varphi=\phi_0-M_{\rm Pl}\frac{G m_{\rm BH}\epsilon_{\rm BH}}{D} e^{-m_s D}.
\end{equation}
Since the BH in SMG has no scalar hair (the scalar field is constant) \cite{PhysRevLett.108.081103}, we have
$\epsilon_{\rm BH}=0$
(Note that, the same result is also obtained by different methods in Appendix \ref{sbh}). Therefore, $|\epsilon_d|=|\epsilon_{\rm NS}-\epsilon_{\rm BH}|=\epsilon_{\rm NS}$ and the constraint becomes
\begin{equation}\label{bound1}
\epsilon_{\rm NS}<6\times10^{-4}\left(\frac{10^4}{N_{\rm GW}}\right)^{\frac14}.
\end{equation}

In SMG, we recall that the screened parameter of a NS can be approximated by \cite{Zhang2017}
\begin{equation}
\epsilon_{\rm NS}=\frac{\phi_0}{M_{\rm Pl}\Phi_{\rm NS}},
\end{equation}
where $\Phi_{\rm NS}=G m_{\rm NS}/R_{\rm NS}$ is the surface gravitational potential of the NS in the NS-BH system.
Then, the upper bound on $\epsilon_{\rm NS}$ can be translated into a bound on the scalar background $\phi_0$ as follows
\begin{equation}\label{phi0}
\frac{\phi_0}{M_{\rm Pl}}<1.2\times10^{-4}\left(\frac{10^4}{N_{\rm GW}}\right)^\frac14\left(\frac{m_{\rm NS}}{1.4 M_\odot}\right)\left(\frac{10~ {\rm km} }{R_{\rm NS}}\right).
\end{equation}
In the previous work \cite{Zhang2017}, we have obtained the constraint $\epsilon_{\rm WD}<4.2\times 10^{-3}$ from the orbital period derivative $\dot{P}$ of the NS-WD system PSR J1738+0333 in SMG. 
And the corresponding constraint on the scalar background is ${\phi_0}/{M_{\rm Pl}}<3.3\times10^{-8}$. This constraint is more tighter than the constraint \eqref{phi0} because the WD is less compact than NS, $\Phi_{\rm WD}/\Phi_{\rm NS}\sim 10^{-4}$. Since the  space-based LISA mission could detect the GW signals of  WD-BH binaries and NS-WD binaries, it is hopeful to  improve this  constraint by the LISA mission.

In the following discussions, we apply the constraint of \eqref{phi0} to some specific SMG models.

\subsection{Chameleon}

The chameleon model is proposed by Khoury and Weltamn \cite{PhysRevLett.93.171104,PhysRevD.69.044026}, which allows the scalar field to evolve on the cosmological time scales while shielding the fifth force by acquiring a large scalar mass in dense energy environment. Since the original chameleon model is ruled out by the combined constraints of the Solar System and cosmology \cite{Zhang2016}, we consider the exponential chameleon model here.
The scalar potential and the conformal coupling function are given by \cite{PhysRevD.70.123518}
\begin{equation}
V(\phi)=\Lambda^4\exp\left(\frac{\Lambda^{\tilde{\alpha}}}{\phi^{\tilde{\alpha}}}\right),\quad A(\phi)=\exp\left(\frac{\tilde{\beta}\phi}{M_{\rm Pl}}\right),
\end{equation}
where $\tilde{\alpha}$ and $\tilde{\beta}$ are the positive dimensionless constants and $\Lambda$ corresponds to the dark energy scale. The scalar background in the host galaxy for a GW event is at the minimum of the effective potential \eqref{veff} , which is given by \cite{Zhang2017},
\begin{equation}
\phi_0=\left(\frac{\tilde{\alpha}M_{\rm Pl}\Lambda^{4+\tilde{\alpha}}}{\tilde{\beta}\rho_b}\right)^{\frac{1}{\tilde{\alpha}+1}}.
\end{equation}
Using the GW constraint \eqref{phi0}, we obtain
\begin{equation}
\frac{\phi_0}{M_{\rm Pl}}=\frac{\Lambda}{M_{\rm Pl}}\left(\frac{\tilde{\alpha}M_{\rm Pl}\Lambda^3}{\tilde{\beta}\rho_b}\right)^{\frac{1}{\tilde{\alpha}+1}}<1.2\times 10^{-4}.
\end{equation}
Substituting the reduced Plank mass $M_{\rm Pl}=2.4\times 10^{18}~ {\rm GeV}$ and the dark energy scale $\Lambda=2.24\times 10^{-3}~{\rm eV}$ into this inequality,
 and assuming that the density of the host galaxy is close to that of the Milky Way $\rho_b=10^{-42}~{\rm GeV}^4$, we obtain the constraint on the parameters of the exponential chameleon model
\begin{equation}
\log_{10}\tilde{\beta}>\log_{10}\tilde{\alpha}-2.8\tilde{\alpha}+0.32~.
\end{equation}

\subsection{Symmetron}

In the symmetron model, the vacuum expectation value of the scalar field depends on the local mass density. In regions of high density, the scalar field is drawn towards $\phi=0$,
 and the effective potential is symmetric under the transformation $\phi \rightarrow -\phi$. In regions of the low density, this symmetry is  broken.
The scalar potential function and the conformal coupling function in this model take the form \cite{PhysRevLett.104.231301}
\begin{equation}
V(\phi)=\mathbb{V}-\frac12\tilde{\mu}\phi^2+\frac{\lambda}{4}\phi^4,\quad A(\phi)=1+\frac{\phi^2}{2M^2},
\end{equation}
where $\tilde{\mu}$ and $M$ are mass scales, $\lambda$ is a positive dimensionless coupling constant, $\mathbb{V}$ is the vacuum energy of the bare potential $V(\phi)$.
Similarly, we obtain the scalar background in the galaxy
$\phi_0={m_s}/{\sqrt{2\lambda}}$
which is proportional to the scalar mass \cite{Zhang2017}.
Assuming the reduced Compton wavelength $m_s^{-1}$ is roughly of the cosmological scales $(m_s^{-1}\sim 1{\rm Mpc})$,
 and using the upper bound ${\phi_0}/{M_{\rm Pl}}<1.2\times10^{-4}$, we have a weak constraint on $\lambda$,
\begin{equation}
\lambda>2.3\times 10^{-107}.
\end{equation}

\subsection{Dilaton}

The dilation model inspired from string theory has an exponential  potential function and a quadratic conformal coupling function \cite{PhysRevD.82.063519}
\begin{equation}
V(\phi)=\mathbb{V}\exp\left(-\frac{\phi}{M_{\rm Pl}}\right), \quad A(\phi)=1+\frac{(\phi-\phi_\star)^2}{2 M^2},
\end{equation}
where $\mathbb{V}$ is a constant with the dimension of the energy density, $M$ labels the energy scale of the theory, and $\phi_\star$ is approximately the value of the scalar field today.

Applying the GW constraint ${\phi_0}/{M_{\rm Pl}}<1.2\times10^{-4}$ to the scalar background
\begin{equation}
\phi_0=\phi_\star+\frac{M^2 \rho_{\Lambda_0}}{M_{\rm Pl}\rho_b},
\end{equation}
we obtain
\begin{equation}
\frac{M}{M_{\rm Pl}}<4.5~.
\end{equation}

\section{Gravitational waves in $f(R)$ gravity}
\label{frgw}

In this section, we consider the GW waveforms in metric $f(R)$ gravity. Since $f(R)$ gravity can be cast into the form of a scalar-tensor theory, we can directly apply the results
of Sec. \ref{waveform} to $f(R)$ gravity. We also obtain the ppE parameters of $f(R)$ gravity and discuss the GW observational constraints on some specific $f(R)$ models.


The total action for $f(R)$ gravity takes the form \cite{RevModPhys.82.451}
\begin{equation}
\label{fr}
S={1\over 16\pi G}\int d^4 x\sqrt{-\bar{g} }\,f(\bar{R})+ S_m[\bar{g}_{\mu\nu},\Psi_m],
\end{equation}
where $\Psi_m$ denotes collectively the matter fields and the overhead bar denotes the quantities in the Jordan frame.
After the field redefinition, $f'(\bar{R})=\exp\left(-\sqrt{\frac{16\pi G}{3}}\phi\right)$,
and the conformal rescaling $g_{\mu\nu}=\exp\left(\frac{-2\phi}{\sqrt{6}M_\text{Pl}}\right)\bar{g}_{\mu\nu}$, this action can be rewritten as Eq. \eqref{action},
with the bare potential 
$V(\phi)=\frac{f'(\bar{R})\bar{R}-f(\bar{R})}{16\pi G f'(\bar{R})^2}$ 
and the conformal coupling function
$A(\phi)=\frac{1}{\sqrt{f'(\bar{R})}}=\exp\left(\frac{\phi}{\sqrt{6}M_\text{Pl}}\right)$ \cite{RevModPhys.82.451,LIU2018286}.


Having rewritten $f(R)$ gravity as a scalar-tensor theory, we can apply the results of Sec. \ref{ppe} to derive the ppE parameters of $f(R)$ gravity and constrain it by GW observations.

Using the relation between $\bar{R}$ and $\phi$, the screened parameter of a NS can be rewritten as
\begin{equation}
\epsilon_{\rm NS}=\frac{\sqrt{6}}{2}\frac{(1-f'(\bar{R}_\infty))}{\Phi_{\rm NS}}.
\end{equation}
where $\bar{R}_\infty=8\pi G \rho_g$ and $\rho_g$ is the average galactic density.

From Eq. \eqref{ppe_smg}, the ppE parameters of  a NS-BH binary system in $f(R)$ gravity are given by,
\begin{align}\label{ppe_nb}
\begin{split}
\alpha_{\rm NS-BH}&=-\frac{5}{256}\frac{\left[1-f'(\bar{R}_\infty)\right]^2}{\Phi_{\rm NS}^2}\eta^{2/5}, \quad \beta_{\rm NS-BH}=-\frac{15}{28672}\frac{[1-f'(\bar{R}_\infty)]^2}{\Phi_{\rm NS}^2}\eta^{2/5},\\
a_{\rm NS-BH}&=-2, \quad ~~~~~~~~~~~~~~~~~~~~~~~~~~~~b_{\rm NS-BH}=-7~.
\end{split}
\end{align}
Similarly, the ppE parameters of a NS-WD binary system in $f(R)$ gravity are
\begin{align}\label{ppe_nb}
\begin{split}
\alpha_{\rm NS-WD}&=-\frac{5}{256}[1-f'(\bar{R}_\infty)]^2\left(\frac{1}{\Phi_{\rm NS}}-\frac{1}{\Phi_{\rm WD}}\right)^2\eta^{2/5}, \quad
 \beta_{\rm NS-WD}=-\frac{15}{28672}[1-f'(\bar{R}_\infty)]^2\left(\frac{1}{\Phi_{\rm NS}}-\frac{1}{\Phi_{\rm WD}}\right)^2\eta^{2/5},\\
a_{\rm NS-WD}&=-2, \quad ~~~~~~~~~~~~~~~~~~~~~~~~~~~~~~~~~~~~~~~~~~~~~~~~~~~b_{\rm NS-WD}=-7~.
\end{split}
\end{align}

 Now, we apply the constraint of Eq.\eqref{ppe_beta} to $f(R)$ gravity. Since this constraint is derived from the potential observations of NS-BH binaries, we should impose it on $\beta_{\rm NS-BH}$, which reads
\begin{equation}\label{1-fr}
\left|1-f'(\bar{R}_\infty)\right|<0.98\times 10^{-4}\left(\frac{10^4}{N_{\rm GW}}\right)^{\frac{1}{4}}\left(\frac{m_{\rm NS}}{1.4 M_\odot}\right)\left(\frac{10{\rm km}}{R_{\rm NS}}\right).
\end{equation}
Note that this constraint is independent of the form of $f(R)$ and  should be satisfied for any $f(R)$ gravity. Let us focus on the specific $f(R)$ models as follows,
\begin{eqnarray}
(A):\; f(R)&=&R- \tilde{m}^2\frac{c_1(R/\tilde{m}^2)^n}{c_2(R/\tilde{m}^2)^n+1},  \;(c_1,c_2,n>0),\label{Hu}\\
(B):\; f(R)&=&R-\tilde{\mu} R_c \tanh\left(\frac{R}{R_c}\right), \; \left(\tilde{\mu},R_c>0\right),\label{Tsu}\\
(C):\; f(R)&=&R-\tilde{\mu} R_c\left[1-\left(1+\frac{R^2}{R_c^2}\right)^{-k}\right], \;(\tilde{\mu},k,R_c>0).\label{Star}
\end{eqnarray}
Model A is proposed by Hu and  Sawicki \cite{Hu:2007nk}, in which the mass scale is $\tilde{m}^2=\frac{8\pi G {\rho}_0}{3}$, where ${\rho}_0 $ is the average matter density in the universe today. Models B and C are proposed by Tsujikawa \cite{PhysRevD.77.023507} and Starobinsky \cite{starobinsky2007disappearing}, respectively, in which $R_c$ roughly corresponds to the order of observed cosmological constant for $\tilde{\mu}=\mathcal{O}(1)$. Since the free parameters of Model A are in one-to-one correspondence with that of Model C \cite{LIU2018286}, we discuss only Models A and B in the following discussions.

In the Hu-Sawicki model, 
the constraint \eqref{1-fr} becomes
\begin{equation}
|1-f'(\bar{R}_0)|<0.98\times 10^{-4}\left(\frac{10^4}{N_{\rm GW}}\right)^{\frac{1}{4}}\left(\frac{m_{\rm NS}}{1.4 M_\odot}\right)\left(\frac{10~{\rm km}}{R_{\rm NS}}\right)\left(\frac{8\pi G \rho_g}{\bar{R}_0}\right)^{n+1}.
\end{equation}
where $\bar{R}_0$ is the scalar curvature of  a spatial flat Friedmann-Lema\^{\i}tre-Robertson-Walker universe at the present epoch \cite{Hu:2007nk}.

In the Tsujikawa model, the constraint \eqref{1-fr} becomes
\begin{equation}
\frac{\tilde{\mu}}{\cosh^2\frac{\tilde{\mu}\bar{R}_\infty}{2\Lambda_0}}<0.98\times 10^{-4}\left(\frac{10^4}{N_{\rm GW}}\right)^{\frac{1}{4}}\left(\frac{m_{\rm NS}}{1.4 M_\odot}\right)\left(\frac{10~{\rm km}}{R_{\rm NS}}\right)\left(\frac{8\pi G \rho_g}{\bar{R}_0}\right)^{n+1}
\end{equation}
where $\Lambda_0$ is the observed cosmological constant \cite{ade2016planck}.
Then,  the inequality
$\frac{\tilde{\mu}}{\cosh^2\frac{\tilde{\mu}\bar{R}_\infty}{2\Lambda_0}}<0.98\times 10^{-4}$
can be satisfied by all $\tilde{\mu}>0$,
where we have used  $\frac{\bar{R}_\infty}{\Lambda_0}=\frac{\rho_g}{\Omega_\Lambda\rho_c}$,  $\Omega_\Lambda=0.692$ and $\rho_c=0.86\times10^{-26}{\rm kg~m^{-3}}$ \cite{ade2016planck} and assumed $\rho_g=10^{-24} {\rm g~cm^{-3}}$.

\section{conclusions}
\label{con}

SMG is a kind of scalar-tensor theories with screening mechanisms to suppress the fifth force in dense regions. Based on the previous work \cite{Zhang2017}, in this paper we have calculated the GW waveforms of an inspiral compact binary system on a quasicircular orbit in general SMG. We find that in SMG there are three propagation degrees, two massless tensor degrees and one massive scalar degree. However, there exist four polarizations in the $E(2)$ classifications, since the massive scalar field induces two polarization modes (the breathing polarization $h_b$ and longitudinal polarization $h_L$). Due to the existence of $h_L$, SMG is   class $II_6$ in the $E(2)$ classification. We have also obtained a simple linear relation between the two scalar polarizations, $h_L=\frac{m_s^2}{\omega_{\rm GW}^2}h_b$, which is a consequence of the linearized scalar field equation and consistent with the previous work \cite{PhysRevD.62.024004}. As a result, the amplitude of the longitudinal mode will decrease with time, which is different from the chirping nature (both amplitude and frequency increase with time) of GWs. Employing the stationary phase approximation, we have derived the Fourier transforms of the four polarization modes, and found a scalar mass induced phase difference $-\frac{m_s^2D}{4\pi f}$ between $\tilde{h}_{L2}$ and $\tilde{h}_+$, which is consistent with the previous results obtained in \cite{PhysRevD.57.2061}. In comparison with the GW waveforms of GR, we have identified the ppE parameters in general SMG. Applying to some specific SMG models, including chameleon, symmetron, dilaton and $f(R)$, the dependences of the ppE parameters on the corresponding model parameters have been obtained. Considering the potential observations of ET on the GWs emitted by NS-BH binaries, we have obtained a constraint of $\epsilon_{\rm NS}<6\times10^{-4}$  even for the conservative estimations with $10^4$ GW events in the redshift range $z<5$, which is a general result and applicable to any SMG model.

\begin{acknowledgements}
We thank Lijing Shao for helpful comments. This work is supported by NSFC No. 11603020, 11633001, 11173021, 11322324, 11653002, 11421303, 11375153, 11675145, 11675143, 11105120, project of Knowledge Innovation Program of Chinese Academy of Science, the Fundamental Research Funds for the Central Universities and the Strategic Priority Research Program of the Chinese Academy of Sciences Grant No. XDB23010200.
\end{acknowledgements}

\appendix

\section{Evaluation of integrals arising in the  waveforms of the scalar polarizations}\label{appendix_a}
We follow the method described in Appendix B of \cite{PhysRevD.85.064041} to calculate the integrals with the Bessel function in Eqs. \eqref{hb} and \eqref{hl}:
\begin{align}
\begin{split}
I_1=&\int_0^\infty dz J_1(z) \frac{1}{u^2}\omega(t-Du)^{\frac13}\cos\left(\Phi(t-Du)\right),\\
I_2=&\int_0^\infty dz J_1(z) \frac{1}{u^3}\omega(t-Du)^{\frac23}\cos\left(2\Phi(t-Du)\right),\\
I_3=&\int_0^\infty dz J_1(z) \left(\frac{1}{u^2}-1\right)\frac{1}{u^2}\omega(t-Du)^{\frac13}\cos\left(\Phi(t-Du)\right),\\
I_4=&\int_0^\infty dz J_1(z) \left(\frac{1}{u^2}-1\right)\frac{1}{u^3}\omega(t-Du)^{\frac23}\cos\left(2\Phi(t-Du)\right),
\end{split}
\end{align}
with $u=\sqrt{1+\big(\frac{z}{m_s D}\big)^2}$ and $\omega(t)=d\Phi(t)/dt$, which cannot be calculated analytically. However, we can obtain
their asymptotic behavior in the wave zone ($D\rightarrow +\infty$) \cite{bender1999advanced,olver1997asymptotics}. Choosing a parameter $\lambda$ such that $m_s D \lambda\gg 1$ and splitting $I_1$ into two parts,
the asymptotic expansion of the first part can be obtained by performing integrations by parts as follows,
\begin{align}
\begin{split}
&\int_0^{m_sD\lambda}dz J_1(z)\frac{1}{u^2}\omega(t-Du)^\frac13 \cos(\Phi(t-Du))\\
=&-J_0(z)\frac{1}{u^2}\omega(t-Du)^\frac13\cos(\Phi(t-Du))\Big|_0^{m_sD\lambda}+\cdots\\
=&\omega(t-D)^\frac13\cos(\Phi(t-D))-J_0(m_sD\lambda)\frac{1}{1+\lambda^2}\omega(t-D\sqrt{1+\lambda^2})^\frac13\cos(\Phi(t-D\sqrt{1+\lambda^2}))+\cdots
\end{split}
\end{align}
where we have used the relation $J_0'(z)=-J_1(z)$.
For the second part, when we perform integrations by parts, it can be exactly canceled with the $\lambda$-dependent terms in the above equation. Therefore, all the contributions that come from the end point  $m_sD\lambda$ can be ignored.

Substituting the asymptotic expression of the Bessel function
\begin{equation}
J_\nu(x)\simeq\sqrt{\frac{2}{\pi x}}\cos\left(x-\frac{\nu\pi}{2}-\frac{\pi}{4}\right),
\end{equation}
into the second part, the integral can be approximated by
\begin{align}
\begin{split}
I_1'&=\int_{m_sD\lambda}^\infty dz\sqrt{\frac{2}{\pi z}}\cos\left(z-\frac{3}{4}\pi\right)\frac{1}{u^2}\omega(t-Du)^\frac13 \cos(\Phi(t-Du))\\
&=\frac12\sqrt{\frac{2m_sD}{\pi}}\int_{\sqrt{1+\lambda^2}}^\infty du\frac{\omega(t-Du)^\frac13}{(u^2-1)^\frac{3}{4}u}\Re \left[e^{i(m_sD\sqrt{u^2-1}-\frac34 \pi+\Phi(t-Du))}+e^{i(m_sD\sqrt{u^2-1}-\frac34 \pi -\Phi(t-Du))}\right],
\end{split}
\end{align}
where $\Re$ denotes the real part of the argument. When $\omega>m_s$, the first term has a stationary point $u_1$ which is determined by
\begin{equation}
\rho'(u_1)=\frac{m_sD u}{\sqrt{u^2-1}}-\omega(t-Du)D\Big|_{u=u_1}=0,
\end{equation}
that is
\begin{equation}
u_1=\frac{\omega(t-Du_1)}{\sqrt{\omega(t-Du_1)^2-m_s^2}},
\end{equation}
where $\rho(u)=m_sD\sqrt{u^2-1}-\frac{3}{4}\pi+\Phi(t-Du)$.

In  real situations, we always have $\omega\gg m_s$. Therefore, the stationary point is very close to unity and we can approximate $u_1$ by
\begin{equation}\label{u1}
u_1=\frac{\omega(t-D)}{\sqrt{\omega(t-D)^2-m_s^2}}.
\end{equation}

Expanding $\rho(u)$ around $u_1$ to the second order
\begin{equation}
\rho(u)=\rho(u_1)+\frac12\rho''(u_1)(u-u_1)^2+\cdots
\end{equation}
then the dominant contribution to the integral $I_1'$ is
\begin{equation}
I_1'\sim \frac12\sqrt{\frac{2m_sD}{\pi}}\frac{\omega(t-Du_1)^\frac13}{(u_1^2-1)^\frac{3}{4}u_1}\Re\Bigg[\sqrt{\frac{2\pi}{\rho''(u_1)}} e^{i(\rho(u_1)+\frac{\pi}{4})}\Bigg].
\end{equation}
Thus, to the leading order, we have  $I_1$
\begin{equation}
I_1\simeq \omega(t-D)^\frac13\cos(\Phi(t-D))-\omega(t-Du_1)^{-\frac23}\sqrt{\omega(t-Du_1)^{2}-m_s^2}\cos\left(\frac{m_s^2D}{\sqrt{\omega(t-Du_1)^{2}-m_s^2}}+\Phi(t-Du_1)\right),
\end{equation}
with $u_1$ being given by Eq. \eqref{u1}.

Similarly, we can obtain the asymptotic expression of the other three integrals
\begin{align}
\begin{split}
I_2 &\sim \omega(t-D)^\frac23\cos(2\Phi(t-D))-\omega(t-Du_2)^{\frac23}\left(1-\frac{m_s^2}{4\omega(t-Du_2)^2}\right)\cos\Bigg(\frac{m_s^2D}{\sqrt{4\omega(t-Du_2)^{2}-m_s^2}}+2\Phi(t-Du_2)\Bigg),\\
I_3 &\sim \frac{m_s^2}{\omega^{\frac83}}\sqrt{\omega^2-m_s^2}\cos\left(\frac{m_s^2D}{\sqrt{\omega^2-m_s^2}}+\Phi\right)\Bigg|_{t-Du_1},\\
I_4 &\sim \frac{m_s^2}{4\omega^2}\left(1-\frac{m_s^2}{4\omega^2}\right)\omega^\frac23\cos\left(\frac{m_s^2D}{\sqrt{\omega^2-m_s^2}}+\Phi\right)\Bigg|_{t-Du_2},\\
\end{split}
\end{align}
where $u_2$ is given by
\begin{equation}
u_2=\frac{2\omega(t-D)}{\sqrt{4\omega(t-D)^2-m_s^2}}.
\end{equation}

\section{The sensitivity of BH in SMG}
\label{sbh}

In Section \ref{ppe}, we derived the screened parameter of BH by the similar way of Appendix A in \cite{1989ApJ...346..366W}. In this Appendix, we will derive this result by a different method.

The BH mass in the Einstein frame $m(\phi)$ is constant and does not evolve with the scalar field \cite{PhysRevLett.83.2699}. From the definition of the sensitivity
\begin{equation}
s\equiv\frac{\partial(\ln m)}{\partial(\ln \phi)}\bigg|_{\phi_0},
\end{equation}
we find that in SMG, both the sensitivity $s_{\rm BH}$ and the screened parameter $\epsilon_{\rm BH}$  of BHs are zero.

The action of the SMG, as a kind of scalar-tensor theories, in the Jordan frame  is
\begin{equation}
S_J=\int d^4x \sqrt{-\bar{g}}\frac{1}{16\pi G}\left[\bar{\phi}\bar{R}-\frac{\omega(\bar{\phi})}{\bar{\phi}}\bar{g}^{\mu\nu}\partial_\mu\bar{\phi}\partial_\nu\bar{\phi}-U(\bar{\phi})\right]+S_m\left[\bar{g}_{\mu\nu},\Psi_m\right],
\end{equation}
where $\omega(\bar{\phi})$ is the coupling function, $U(\bar{\phi})$ is the scalar potential and $\bar{R}$ is the Ricci scalar derived from the Jordan frame metric $\bar{g}_{\mu\nu}\equiv A(\phi)^2g_{\mu\nu}$.
The Jordan frame Ricci scalar $\bar{R}$ and the Einstein frame Ricci scalar $R$ are related by the relation $\bar{R}=A^{-2}[R-6g^{\mu\nu}\nabla_\mu\nabla_\nu\ln A-6g^{\mu\nu}(\nabla_\mu\ln A)\nabla_\nu\ln A]$ \cite{wald1984general}. Using this relation, we
obtain
\begin{align}
\omega(\bar{\phi})=2\pi G\left(\frac{d\ln A(\phi)}{d \phi}\right)^{-2}-\frac32,~~
U(\bar{\phi})=\frac{V(\phi)}{A(\phi)^4},~~
\bar{\phi}=A(\phi)^{-2}.
\end{align}
The BH mass in the Jordan frame is given by $\bar{m}(\bar{\phi})=A(\phi)^{-1}m$ \cite{PhysRevLett.83.2699}, that is,  $\bar{m}(\bar{\phi})=\bar{\phi}^\frac12 m$. Thus, the sensitivity of a BH in the Jordan frame is
\begin{equation}
\bar{s}_{\rm BH}=\frac{\partial(\ln \bar{m})}{\partial(\ln \bar{\phi})}\bigg|_{\bar{\phi}_0}=\frac12.
\end{equation}
So, we find that the sensitivity of a BH in the general SMG is the same as that in Brans-Dicke theory \cite{1975ApJ...196L..59E,1989ApJ...346..366W}. In addition,
Sotiriou and Faraoni  proved that a stationary BH in a general scalar-tensor theory is the same as in GR and that the scalar field is constant in this spacetime \cite{PhysRevLett.108.081103}.
As a result, the screened parameter of a BH, which is zero, also satisfies the relation
\begin{equation}
\epsilon_a=\frac{\phi_0-\phi_a}{M_{\rm Pl}\Phi_a},
\end{equation}
although this relation is derived from a star composed of a perfect fluid \cite{Zhang2016}.

%

\end{document}